\documentclass[preprint,preprintnumbers,amsmath,amssymb]{revtex4}

\usepackage{graphicx}% Include figure files
\usepackage{dcolumn}% Align table columns on decimal point
\usepackage{bm}% bold math

\begin{document}

%................................................
%\begin{figure}[htb]
%\centerline{\includegraphics[scale=0.2]{openwebmail.eps}}
%\caption{IPICyT}
%\end{figure}
%....................................................

\setcounter{MaxMatrixCols}{10}

\title{Cellular Automata Modeling of Continuous Stirred Tank Reactors}

\author{J.E. P\'erez-Terrazas$^1$\footnote{E-mail: jenrique@ipicyt.edu.mx},
V. Ibarra-Junquera$^2$\footnote{E-mail: vrani@ipicyt.edu.mx} \& H.C.
Rosu$^1$\footnote{E-mail: hcr@ipicyt.edu.mx, Corresponding author}}

\affiliation{$^1$ Potosinian Institute of Science and Technology,
Apartado Postal 3-74 Tangamanga, 78231 San Luis Potos\'{\i}, Mexico\\
$^2$ Faculty of Chemical Sciences, University of Colima,
Coquimatl\'an, Col., Mexico}

\date{Sept. 24, 2006}  % It is always \today, today,

%\centerline{ArXiv: }

%\centerline{code}

%,,,,,,,,,,,,,,,,,,,,,,,,,,,,,,,,,,,,,,,,,,,,,,,,,,,,,,,,,,,,,,,,,,,,,,,,,,,,,
\begin{abstract}
The classical dynamical systems model of continuous stirred tank
reactors (CSTR) in which a first order chemical reaction
  takes place is reformulated in terms of stochastic cellular automata by extending previous
  works of Seyborg \cite{Seybold 1997} and Neuforth \cite{Neuforth 2000} by including the feed flow of chemical reactants.
  We show that this cellular
  automata procedure is able to simulate the dilution rate and the mixing process in the CSTR, as well as the details of the heat removal
  due to the jacket. The cellular automata approach is expected to be of
considerable applicability at any industrial scales and especially
for any type of microchemical systems.

\medskip

 %Chemical kinetics and dynamics,

%82.40.Ck Pattern formation in reactions with diffusion, flow and
%heat transfer (see also 47.54.-r Pattern selection; pattern
%formation and 47.32.C- Vortex dynamics in fluid dynamics)

%47.70.Fw Chemically reactive flows
%02.70.-c Computational techniques; simulations

\end{abstract}
%,,,,,,,,,,,,,,,,,,,,,,,,,,,,,,,,,,,,,,,,,,,,,,,,,,,,,,,,,,,,,,,,,,,,,,,,,,,,,,

\vspace*{10pt}
%\keywords{The contents of the keywords}

%{\bf PACS} numbers: 05.45.-a, 87.10.+e}  % PACS, the Physics and Astronomy
%Keywords: barotropic cosmology

\maketitle

\section{Introduction}

A chemical reactor could be any vessel containing chemical
reactions. In general, a reactor is designed such as to maximize the
yield of some particular products while requiring the least amount
of money to purchase and operate. Normal operating expenses include
energy input, energy removal, raw material costs, labor, etc. Energy
changes can occur in the form of heating or cooling, or agitation.
The latter is quite important because an appropriate mixing has a
large influence on the yield. Therefore, the design and operation of
mixing devices often determines the profitability of the whole
plant.Theoretically, the effect of stirring in reactant media have
also attracted considerable attention \cite{theory}.%\\

In particular, in the widely developed continuous stirred tank
reactors (CSTR) one or more fluid reagents are introduced into a
tank equipped with an impeller while the reactor effluent is removed
\cite{Epstein}. The impeller stirs the reagents to ensure proper
mixing. Classical CSTR dynamical models, based on coupled
deterministic ordinary differential equations (ODEs), are the usual
approach to chemical systems at the macroscopic scale. It has been
demonstrated to have considerable usefulness. However, their
validity relies on many assumptions that limit the situations in
which they can be applied. One of the most important is that
chemical systems are discrete at the molecular level and statistical
fluctuations in concentration and temperature occur at the local
scale. The elaboration of models considering this discreteness is
important. Seyborg \cite{Seybold 1997} and Neuforth \cite{Neuforth
2000} have shown that stochastic cellular automata models can be
successfully applied in simulating first order chemical reactions.
In their papers, they worked on a squared arrangement of cells, each
of them having a chemical reactive. The reactions are performed by
considering a probability of change, from reactive A to reactive B,
proportional to the kinetics constant that defines the chemical
equation. However, this type of calculation does not apply directly
to the CSTR case, where a chemical feed flow is present.

 In this paper we extend the stochastic CA model to CSTRs by
simulating the feed flow flux by means of a random selection of a
subset of cells to which the flux conditions with respect to
chemical concentration and temperature are imposed. We would like to
remark that mixing in a stirred tank is complicated and not well
described despite the extensive usage of dimensionless numbers and
models based on ODEs \cite{Chakraborty 2003}. Therefore, more
accurate models are essential for developing and testing control
strategies or even to explore new reactor geometries. The
organization of the paper is the following. Section \ref{S2}
presents shortly the standard ODE-based CSTR model. Section \ref{S3}
describes the CA method that we implemented for jacketed CSTRs. The
simulations are displayed and briefly discussed in Section \ref{S4}.
The paper ends up with several concluding remarks.

%>>>>>>>>>>>>>>>>>>>>>>>>>>>>>>>>>>>>>>>>>>>>>>>>>>
\section{The Deterministic Dynamical Systems Model}\label{S2}

As already mentioned, we consider an ideal jacketed CSTR where the
following exothermic and irreversible first-order reaction is taking
place:
%....................
\begin{eqnarray}
    A \longrightarrow B \nonumber
\end{eqnarray}

The CSTR modeling equations in dimensionless form are the following
\cite{Silva 1999}
\begin{eqnarray}
    \frac{d\,X_1}{d\,\tau} &=& -\phi\,X_1\,k(X_2)+q\,(X_{1_{f}}-X_1)\\
    \frac{d\,X_2}{d\,\tau} &=& \beta\,\phi\,X_1\,k(X_2)-(q+\delta)\,X_{2}+\delta\,X_3+q\,X_{2_{f}}\\
    \frac{d\,X_3}{d\,\tau} &=&
    \frac{q_c}{\delta_1}\,\left(X_{3_{f}}-X_3\right)+\frac{\delta}{\delta_1\,\delta_2}\,(X_2-X_3)~,
\end{eqnarray}
%............................
where $X_1$, $X_2$, and $X_3$ are the dimensionless concentration,
temperature, and cooling jacket temperature, respectively. We note
that it is possible to use the dimensionless coolant flow rate,
$q_c$, to manipulate $X_2$.

\begin{figure}[htb]
        \begin{center}
            \includegraphics[height=5cm]{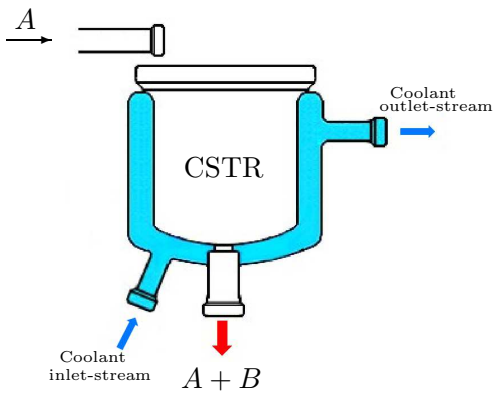}
                \put(-109,-10){$A+B$}
                \put(-160,122){\line(-1,0){15}} \put(-160,122){\vector(1,0){2}}
                \put(-172,126){$A$}
                \put(-155,0){\tiny{Coolant}}
                \put(-159,-7){\tiny{inlet-stream}}
                \put(-30,100){\tiny{Coolant}}
                \put(-34,95){\tiny{outlet-stream}}
                \put(-108,70){CSTR}
        \end{center}
\caption{Schematic representation of the jacketed CSTR.}
\end{figure}
The relationships between the dimensionless parameters and variables
and the physical variables are the following:
\begin{eqnarray}
    k(X_2)=\exp\left(\frac{X_2}{1+X_2\,\gamma^{-1}}\right),\ \gamma=\frac{E}{R\,T_{f_{0}}},\ X_3=\frac{T_c-T_{f_{0}}}{T_{f_{0}}}\,\gamma,\ X_2=\frac{T-T_{f_{0}}}{T_{f_{0}}}\,\gamma\nonumber\\
    \beta=\frac{(-\Delta H)\ C_f}{\rho C_p t_{f_0}},\
    \delta=\frac{U\ A}{\rho \ C_p\ Q_0},\ \phi=\frac{V}{Q_0}k_0\
    e^{-r}, \ X_1=\frac{C}{C_{f_0}},\ X_{1_f}=\frac{C_f}{C_{f_0}},\
    \nonumber\\
    X_{2_f}=\frac{T_f-T_{f_0}}{T_{f_0}}\gamma,\
    \delta_1=\frac{V_c}{V},\ \tau=\frac{Q_0}{V}t,\
    X_{3_f}=\frac{T_{c_f}-T_{f_0}}{T_{f_0}}\gamma,\
    \delta_2=\frac{\rho_c\ C_{p_c}}{\rho\ C_p} \nonumber
\end{eqnarray}
where the meaning of the symbols is given in Table 1. In the rest of
the paper we shall use the solution of this ODE model as the
theoretical case with which the CA simulations will be compared.

\begin{table}[htb]
\centering \caption{Parameters of the model} \label{tab:1}
\begin{tabular}{lll}
\hline\noalign{\smallskip}
Symbol & \ \ \ \ \ \ \ \ \ \ \ Meaning & \ \ \ \ Value   \\
            & \ \ \ \ \ \ \ \ \ \             & (arb. units) \\
\noalign{\smallskip}\hline\noalign{\smallskip}
$C$ & Reactor composition&\ \ \ \ \ $0.001$   \\
$C_f$ & Feed composition&\ \ \ \ \ $1.0$  \\
$q$ & Dimensionless reactor feed flow rate&\ \ \ \ \  $1.0$  \\
$q_c$ & Dimensionless coolant flow rate&\ \ \ \ \ $1.65$  \\
$q_{cs}$ & Steady-state value of $Q$ &\ \ \ \ \ $1.0$  \\
$T$ & Reactor temperature &\  \ \ \ \ $1.0$  \\
$T_c$ & Coolant temperature &\ \ \ \ \ $1.0$  \\
$UA$ & Heat transfer coefficient times the heat transfer area &\ \ \ \ \ $1.0$ \\
$V$ & Reactor volume &\ \ \ \ \ $1.0$  \\
$V_c$ & Cooling jacket volume &\ \ \ \ \ $1.0$  \\
$X_{1f}$ & Dimensionless feed concentration &\ \ \ \ \ $1.0$  \\
$X_{2f}$ & Dimensionless feed temperature &\ \ \ \ \ $0.0$  \\
$X_{3f}$ & Dimensionless coolant feed temperature&\ \ \ \ \ $1.0$  \\
$\beta$ & Dimensionless heat of reaction &\ \ \ \ \ $8.0$  \\
$\delta _1$ & Dimensionless volume ratio &\ \ \ \ \ $0.1$  \\
$\delta _2$ & Dimensionless density multiplied by the heat capacity of coolant &\ \ \ \ \ $1.0$  \\
$\phi$ & Hill's threshold parameter &\ \ \ \ \ $0.072$  \\
$\gamma$ & Dimensionless activation energy &\ \ \ \ \ $20.0$  \\
$\rho _c C_{p_c}$ & Density multiplied by the heat
capacity of coolant&\ \ \ \ \ $1.0$  \\
$\tau$ & Dimensionless time &\ \ \ \ \ $--$  \\
\\\noalign{\smallskip}\hline
\end{tabular}
\end{table}

%$C$   Reactor composition, $C_f$  Feed composition, $q$
%Dimensionless reactor feed flow rate, $q_c$  Dimensionless coolant
%flow rate, $q_{cs}$ Steady-state value of $Q$,   Feed flow rate
%$Q_c$ Coolant flow rate, $T$   Reactor temperature, $T_c$ Coolant
%temperature, $UA$ Heat transfer coefficient multiplied by the heat
%transfer area, $V$ Reactor volume, $V_c$ Cooling jacket volume,
%$X_{1f}$ Dimensionless feed concentration, $X_{2f}$ Dimensionless
%feed temperature, $X_{3f}$ Dimensionless coolant feed temperature,
%$\beta$ Dimensionless heat of reaction, $\delta$ Dimensionless heat
%transfer coefficient, $\delta_1$ Dimensionless volume ratio,
%$\delta_2$ Dimensionless density multiplied by the heat capacity
%ratio, $\phi$ Damk\"{o}hler number (at nominal flow rate), $\gamma$
%Dimensionless activation energy. $\rho C_p$ Density multiplied by
%the heat capacity, $\rho_c C_{p_c}$ Density multiplied by the heat
%capacity of coolant, $\tau$ Dimensionless time.
The nominal parameter values used here are given by:
\begin{eqnarray}
\beta = 8.0,\ \delta = 0.3,\ X_{1_f} = 1.0,\ X_{3_f} = 1.0,\ q =
1.0,\ q_{c_s} = 1.65\nonumber\\ \gamma = 20.0,\ X_{2_f} = 0.0,\ \phi
= 0.072,\ \delta_1 = 0.1,\ \delta_2 = 0.5~.\nonumber
\end{eqnarray}

\section{Stochastic CA Model for Jacketed CSTR}\label{S3}

The process simulated in this work is the exothermic reaction that
converts a chemical A into a product B in a jacketed CSTR. Our model
is composed of three squared arrangements of cells, all of the same
size. The first lattice is for chemicals A and B and represents the
chemical distribution in the tank reactor. In each cell there is
only one unit of reactive A or one unit of product B (not necessary
representing a single molecule), under the condition that all cells
are occupied. The second arrangement is for the tank temperature. It
contains the temperatures $t_{ij}$ in real values, with each cell in
this arrangement corresponding to the respective cell in the first
arrangement. The third arrangement represents the coolant system. We
have used a squared lattice of the same size as the temperature
array in such a way that each temperature tank cell is in
``contact'' with a coolant jacket cell.

In our model the first process in each time step is the irreversible
conversion of chemical A into product B. The conversion rate is
determined by $\phi\,k(X_{2})$  as in \cite{Silva 1999}, where
$X_{2}$ is the average temperature of the tank temperature
arrangement. This first order kinetics
 ``constant'' is multiplied by the time step in order to get the proportion
of the reactive A that is expected to be converted into product B in
each evolution step. This number could also be considered as the
probability that a molecule of chemical A would be converted in
product B if the time step is small enough. Such proportion is
compared with a randomly generated number, one different random
number for each cell in the arrangement containing reactive A. If
the random number is less than the proportion, the reactive A is
changed for product B in the cell. Since the reaction is exothermic,
the temperature value in the
temperature array is increased by $\beta$ (according to Eq.~(2)) in the corresponding cell.\\

The second simulated process in our model is the tank temperature
diffusion that can be simultaneously considered as an energy
diffusion. It can be performed by means of finite differences, but
in order to obtain a model almost ODE independent we have
implemented a moving average method, where the value of the
temperature in a cell at the next time step is the average
temperature of its neighborhood. This procedure gives similar
results to those of finite differences, as shown by Weimar for
reaction-diffusion systems simulated by cellular automata
\cite{Weimar 1997}. We used a square neighborhood formed by
$(2R+1)^{2}$ cells, where $R$ is the number of steps that we have to
walk from the center of the cell in order to reach the most far
horizontal (vertical) cell in the neighborhood.

The third simulated process is the tank feed flow. We have simulated
the feed flow rate $q$ in a stochastic way. In order to get an
approximation to the proportion of the tank that must be replaced by
the incoming flow, $q$ is multiplied by the time step and by the
total number of cells in the arrangement. This give us a real number
$x$. Then, following Weimar \cite{Weimar 1997}, we used a
probabilistic minimal noise rule, \textit{i.e.} it is defined the
probability $p = x- \lfloor x \rfloor$ in order to decide if
$\lfloor x \rfloor$ or $\lfloor x \rfloor +1$ cells will be replaced
by the flow. We choose $\lfloor x \rfloor$ with probability $1-p$
and $\lfloor x \rfloor +1$ with probability $p$. This method
conserves the proportion $x$ in a statistical way. Subsequently, a
cell is selected in a random way, by means of two random numbers
which are used to select a row and a column, in such a way that all
cells has the same probability of being selected. If the cell has
been selected in the same time step, a new selection is made. This
is repeated until we have reached the number of cells that must be
replaced. Finally, the selected cells are changed in the temperature
arrangement by the feed flow temperature $X_{2f}$, and in the
reactive arrangement it is put a unit of reactive A with probability
$X_{1f}$, that represent the concentration of chemical A in the feed
flow. In the simulations presented in this work we used $X_{1f}=1$.
This method of flow simulation could be improved in several ways, in
order to simulate different tank geometries or for showing the flow
direction. However, in this work we want only to show that the CA
method could fit the CSTR behavior in a very good approximation,
with the advantage of spatial analysis.

The fourth simulated process is the energy interchange between the
tank and the jacket. This has been done by directly calculating the
energy/temperature interchange between each tank temperature cell
and its corresponding jacket temperature cell. This interchange is
dictated by the difference between the two temperatures and it is
weighted by $\delta$, $\delta_{1}$, and $\delta_{2}$ as in Eqs.~(1)
and (2). The fifth and the sixth simulated processes are the coolant
flow and the coolant temperature diffusion, respectively. Both of
them are performed in a similar way as for the concentration and
temperature tank.

\section{Simulations}\label{S4}

\begin{figure}[htb]
\centering
\includegraphics[scale=0.5]{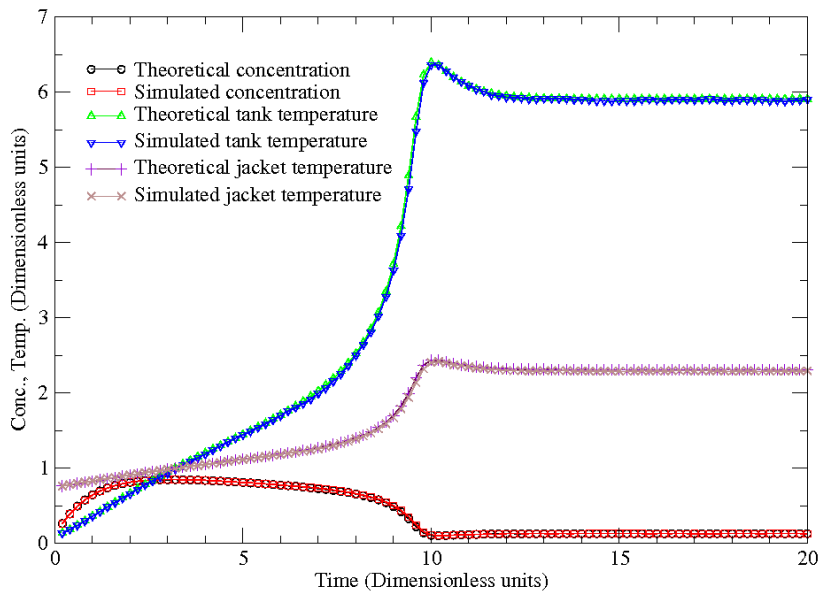}
\caption[]{Comparison between the curves from the differential
equations and the curves obtained from simulations with the cellular
automata approach. Initial values are: $X_{1}=0.1$, $X_{2}=0.1$,
$X_{3}=0.1$. For the six curves: 100 points separated by a
dimensionless time of 0.2 were taken from 20000-point simulations
with a dimensionless time step of 0.001.}
 %\textit{i.e.} 20000 points.}
 \label{comparativo}
\end{figure}

In this section we first present the comparison between the curves
obtained by differential equations and those obtained from the
implemented cellular automata model (see Fig.~\ref{comparativo}). It
is clear that cellular automata simulations resemble with excellent
agreement the values for the concentration of chemical A, the tank
temperature and the jacket temperature at all times. Using a time
step of 0.001 it is shown that the curves coincide at the initial
time, transient time and for stable state. We have found that we can
maintain this remarkable fitness by properly adjusting the time step
to a sufficiently small value.
%However, for a time step greater than 0.001 the peaks
%appear delayed with respect to those obtained from the differential
%equations. This can be explained by the fact that in the CA approach
%the reaction, temperature diffusion, and energy interchange
%processes are performed in a serial way instead of simultaneously,
%as in differential
%equation evaluation.\\

%................FIG. 2

\begin{figure}[htb]
\centering
\includegraphics[scale=0.55]{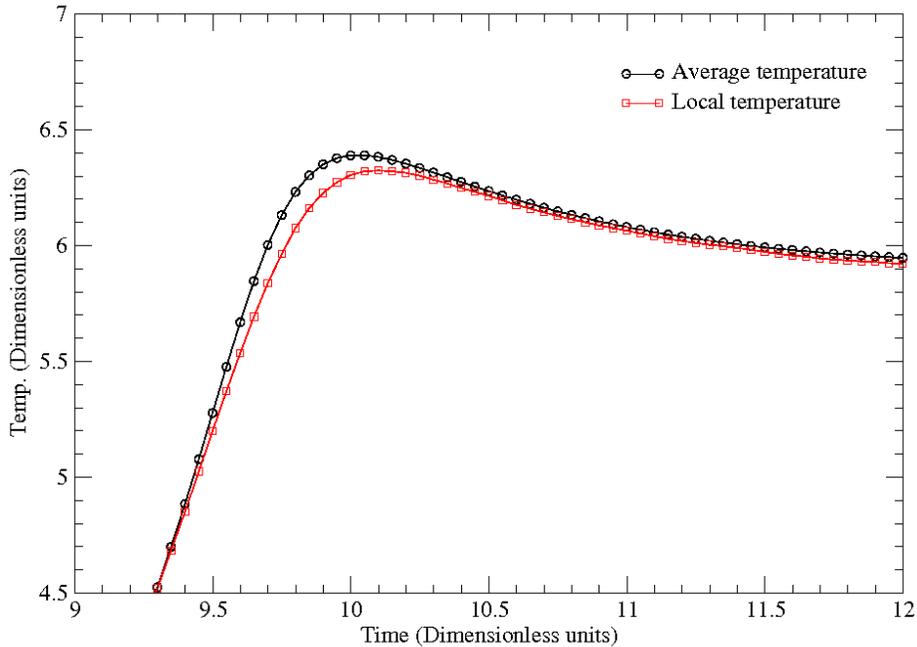}
\caption[]{Comparison between the tank temperature evolution curves
for a kinetic constant based on average temperature and for kinetics
constant calculated on the base of the local temperature. Initial
values are: $X_{1}=0.1$, $X_{2}=0.1$, $X_{3}=0.1$. The array is of
$1000 \times 1000$ cells and the dimensionless time step is of
0.001; one temperature diffusion step per time step.} \label{local}
\end{figure}

When a kinetic constant based on the average temperature is used, it
is implicity assumed that the mixing in the CSTR is perfect,
vanishing any temperature inhomogeneities. One could ask what could
be the change in the tank behavior if the mixing is almost perfect.
It could be studied by using a model that consider the spatial
distribution of temperature. We studied this effect by calculating
the kinetic constant for each cell based in the its correspondent
temperature. The effect for a $1000 \times 1000$ cells array and a
time step of 0.001 is shown in Fig~\ref{local}. It could be observed
that the tank temperature curve for the perfect mixing and the one
for the locally calculated kinetic constant are the same for almost
all times. However, they separate during the transient period,
leading to a reduction in the magnitude of the peak and a little
delay in its appearance. The curve was calculated with a tank
temperature diffusion process per time step with a $R=1$
neighborhood. If more diffusion steps are used per time step the
curves obtained tend to the theoretical one as is expected. Besides,
it could be also interpreted as the effect of a perfect mixed tank,
but with material
where each component tends to maintain its energy.\\

%[Aqui hacen falta al menos dos correcciones. El pico al final del
%transient time tiene nombre y no me lo se. Y segundo: esto ultimo de
%material que no le gusta compartir su energia, si existe este tipo
%de material? habria que dar un ejemplo]

%.....................FIG. 3

The ODEs generally represent the characteristics of the global
system, where it has enough number of elements such that the
statistical fluctuations are of small amplitude. %until they are not important in
%the global behavior.
However, when the system size and the quantity of elements are
diminished, the statistical fluctuations could be of increasing
importance. In this way, another advantages of the cellular automata
method proposed here is its flexibility with respect to the reactor
size, and its stochastic nature,
that allows to study how much the system could be affected by the initial conditions and by the
stochastic features of the process.\\

We have performed several simulations applying the parameters
presented above in arrangements of small size, where the model is no
longer representative for a 3-dimensional CSTR at such scales.
However, we can think of it as a representation of a catalytic
surface dividing two regions, one carrying the chemical A and the
other as a temperature reservoir. Therefore, this model is a simple
approach, useful as a first approximation, in the analysis and study
of microreactors or even nanoreactors. We recall that the usage of
microreactors for {\em in situ} and on-demand chemical production is
gaining increasing importance as the field of microreaction
engineering has already demonstrated potential to impact a wide
spectrum of chemical, biological, and process system applications
\cite{Jensen}. There are already many successfully developed
microreactors for chemical applications such as partial
oxidation reactions \cite{partial}, %of ammonia \cite{ammonia},
%nitration \cite{nitration},
phosgene synthesis \cite{phosgene}, multiphase processing
\cite{multiphase}, and (bio)chemical detection \cite{detection}.

%[Aqui van unas palabras de Vrani acerca de bioreactores]\\

\begin{figure}[htb]
\centering
\includegraphics[scale=0.55]{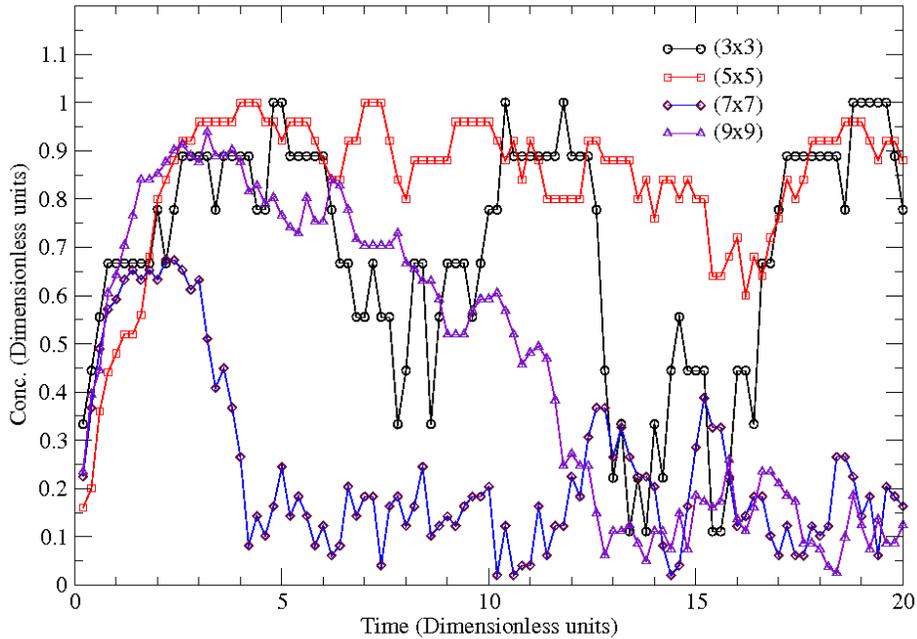}
\caption[]{Behaviors that could be found in systems with small
number of elements (cells for the CAs and clusters of molecules in
the real case). These behaviors are different from that expected in
systems with a large number of elements. The initial values are
$X_{1}=0.1$, $X_{2}=0.1$, $X_{3}=0.1$. The employed time step is
0.001. One hundred points separated by a time lag of 0.2 were taken
from 20000-point simulations with a time step of 0.001;
%\textit{i.e.} 20000 points.
one temperature diffusion step per time step.} \label{smallconc}
\end{figure}

Figure \ref{smallconc} displays the variability that could be found
in CSTR systems at small scales. It is clear evidence that the
statistical fluctuations are a primordial issue at this scale. In
addition, one could notice that the dynamical behavior could be
totally different to the expected behavior of a larger system e.g.,
$1000 \times 1000$ cells arrangement.

%.................FIG. 4

\begin{figure}[htb]
\centering
\includegraphics[scale=0.55]{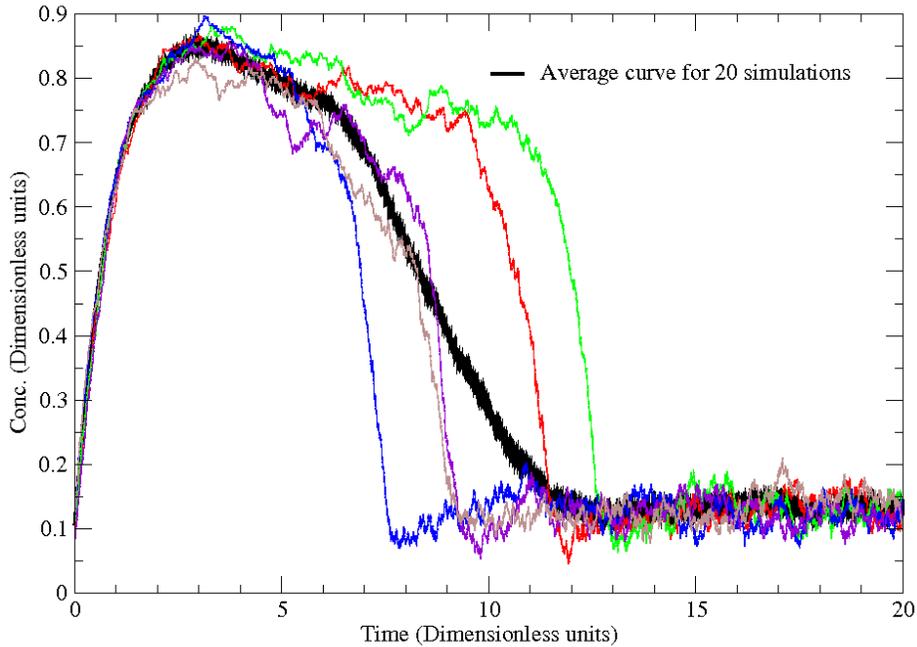}
\caption[]{Different concentration behaviors of chemical A for
systems of the same size ($20 \times 20$) that are treated by the
same method and could be the underlying dynamic characteristic of
microreactors. Five simulations are displayed together with the
average curve (thick line) of 20 individual simulations. The initial
configuration and the stochasticity introduced in the model lead to
a time-distributed behavior. The initial values are: $X_{1}=0.1$,
$X_{2}=0.1$, $X_{3}=0.1$. The time step is 0.001; one temperature
diffusion step per time step.} \label{rep}
\end{figure}

Finally, the study of small systems by direct simulation using
stochastic simulations could give us insight in how an open CSTR
system could behave when the statistical fluctuations and the
initial configuration are important. In Fig.~\ref{rep} one can see
that the possible behaviors of a system of size $20 \times 20$ have
big deviations from the average (theoretical) value. This kind of
variability is not provided by pure ODEs (without a stochastic
term). We think that this stochastic CA approach could be an
important tool for testing control strategies since the CA approach
could be seen as a step between the ODEs models and the specific
experimental situation.

%....................FIG. 5

%falta ver si funciona lo de poner grumos en la alimentacion del tanque

\section{Concluding Remarks}\label{S5}

A cellular automata approach for the CSTR with cooling jacket has
been presented in this paper. It is able to reproduce the CSTR
dynamical behavior calculated by ODE's with a good approximation and
in an easy way. The stochastic model presented allow us to study
what could be the behavior of the variables of the tank when the
reaction probability depends on the local temperature. It also give
us an approach to study systems with few elements, as could be micro
and nanoreactors, as could be catalytic membranes separating two
phases. The main advantages of the CA approach presented here are
its stochastic nature and the direct involvement of a spatial
structure. This also represents a tool for studying the role of
initial configuration and stochastic fluctuations in systems with
few elements. Additionally, the CA approach is a clear improvement
of the CSTR modeling and moreover can be applied to different
reactor and jacket geometries, as well as for considering in more
detail the real mass flow in the tank reactor-geometry.

%\newpage

\end{document}